\documentclass[
a4paper, 
reprint, 
showkeys, 
nofootinbib,
superscriptaddress, 
twoside,
aps,
prl,
nobibnotes,
longbibliography]{revtex4-1}
\usepackage[english]{babel}
\usepackage[utf8]{inputenc}
\usepackage{amsthm}
\usepackage{mathtools}
\usepackage{physics}
\usepackage{xcolor}
\usepackage{graphicx}
\usepackage[left=23mm,right=13mm,top=35mm,columnsep=15pt]{geometry} 
\usepackage{adjustbox}
\usepackage{placeins}
\usepackage[T1]{fontenc}
\usepackage{lipsum}
\usepackage{csquotes}
\usepackage{hyperref}
\usepackage{soul}
\usepackage{color}
\usepackage{amsmath}
\usepackage{amsfonts}
\usepackage{amssymb}
\usepackage{upgreek}
\usepackage{changes}
\usepackage{comment}
\AtBeginDocument{
    \newwrite\bibnotes
    \def\bibnotesext{Notes.bib}
    \immediate\openout\bibnotes=\jobname\bibnotesext
    \immediate\write\bibnotes{@CONTROL{REVTEX41Control}}
    \immediate\write\bibnotes{@CONTROL{
    apsrev41Control,author="08",editor="1",pages="1",title="0",year="1"}}
     \if@filesw
     \immediate\write\@auxout{\string\citation{apsrev41Control}}
    \fi
}

\def\ii{{\rm i}}  \def\ee{{\rm e}}

        \def\Eb{{\bf E}}                \def\Jb{{\bf J}}      \def\kb{{\bf k}}        \def\pb{{\bf p}}  \def\Qb{{\bf Q}}      \def\rb{{\bf r}}       
    \def\zz{\hat{\bf z}}          \def\RR{\hat{\bf R}}

\begin{document}

\title{Exploring single-photon recoil on free electrons}
\author{Alexander Preimesberger}
\affiliation{Vienna Center for Quantum Science and Technology, Atominstitut, TU Wien, Vienna, Austria}
\affiliation{University Service Centre for Transmission Electron Microscopy,TU Wien, Wiedner Hauptstraße 8-10/E057-02, 1040 Wien, Austria}

\author{Dominik Hornof}
\affiliation{Vienna Center for Quantum Science and Technology, Atominstitut, TU Wien, Vienna, Austria}
\affiliation{University Service Centre for Transmission Electron Microscopy,TU Wien, Wiedner Hauptstraße 8-10/E057-02, 1040 Wien, Austria}

\author{Theo Dorfner}
\affiliation{Vienna Center for Quantum Science and Technology, Atominstitut, TU Wien, Vienna, Austria}
\affiliation{University Service Centre for Transmission Electron Microscopy,TU Wien, Wiedner Hauptstraße 8-10/E057-02, 1040 Wien, Austria}

\author{Thomas Schachinger}
\affiliation{University Service Centre for Transmission Electron Microscopy,TU Wien, Wiedner Hauptstraße 8-10/E057-02, 1040 Wien, Austria}

\author{Martin Hrto\v{n}}
\affiliation{Institute of Physical Engineering, Brno University of Technology, 616 69 Brno, Czech Republic}
\affiliation{Central European Institute of Technology, Brno University of Technology, 612 00 Brno, Czech Republic}

\author{Andrea Kone\v{c}n\'{a}}
\affiliation{Institute of Physical Engineering, Brno University of Technology, 616 69 Brno, Czech Republic}
\affiliation{Central European Institute of Technology, Brno University of Technology, 612 00 Brno, Czech Republic}

\author{Philipp Haslinger}
\email[]{philipp.haslinger@tuwien.ac.at}
\affiliation{Vienna Center for Quantum Science and Technology, Atominstitut, TU Wien, Vienna, Austria}
\affiliation{University Service Centre for Transmission Electron Microscopy,TU Wien, Wiedner Hauptstraße 8-10/E057-02, 1040 Wien, Austria}

\date{\today}

\begin{abstract}
Recent advancements in time-resolved electron and photon detection enable novel correlative measurements of electrons and their associated cathodoluminescence (CL) photons within a transmission electron microscope. These studies are pivotal for understanding the underlying physics in coherent CL processes. We present experimental investigations of energy-momentum conservation and the corresponding dispersion relation on the single particle level, achieved through coincidence detection of electron-photon pairs. This not only enables unprecedented clarity in detecting weak signals otherwise obscured by non-radiative processes but also provides a new experimental pathway to investigate momentum-position correlations to explore entanglement within electron-photon pairs.

\end{abstract}


\maketitle

Transmission electron microscopes (TEMs) are exceptional tools for investigating samples down to single-atom resolution \cite{Reimer2008,batson2002sub}. Standard TEM imaging relies on detecting energetic (30--300 keV) electrons of the primary beam transmitted through a thin sample, where most of the electrons interact with the sample elastically. However, a small portion of the inelastically scattered primary electrons can produce the emission of secondary electrons or photons from the sample. Spectral analysis of emitted X-ray photons is often used for extracting elemental composition \cite{D'Alfonso_EDX_atomic_2010, Chu_EDX_atomic_2010}, while detection of low-energy (typically visible and near-infrared) photons can be linked to the local optical and electronic properties \cite{polman2019CL_electron}.
The electrons can bring the sample into an excited state, from which it can consequently de-excite along a variety of competing pathways. Optical photons emitted during such a de-excitation cascade are traditionally labeled as cathodoluminescence (CL) \cite{Kociak2017,dangCathodoluminescenceNanoscopyState2023} and carry information on radiative transitions between sample energy levels. CL is often observed in semiconductor and excitonic nanostructures, such as thin films \cite{vu2022exciton}, quantum dots and wells \cite{grundmannUltranarrowLuminescenceLines1995,Meuret2018}, or in crystal defects, including color centers \cite{Meuret2015,sauerRevisedFineSplitting2000,angell2024_SV_colorCenter_CL,costantini2017cL_colorCenters_JEM-1300NEF}.

An electron can also trigger a direct emission of a photon, which is often denoted as ``coherent'' CL. Such photons can propagate within the sample material (e.g., the so-called Cherenkov radiation \cite{Cherenkov1937,
stoger-pollachCerenkovLossesLimit2006,Scheucher2022}) or radiate into the vacuum surrounding the sample. In the latter case, we distinguish the Smith-Purcell radiation produced when electrons interact with a periodic structure \cite{Smith1953,KaminerSmithPurcell2017}, the transition radiation emerging when electrons traverse material boundaries \cite{Yamamoto2001,Brenny2016_TR,Stoger-Pollach2017,Chen2023_TR}, and photons produced due to the excitation of localized electromagnetic modes, including Mie modes in dielectric and localized surface plasmons in metallic nanostructures \cite{Vesseur2007,kuttgeLocalDensityStates2009}. Importantly, photons generated during the aforementioned processes carry energy and momentum related to the electron energy loss and momentum recoil of the fast electron \cite{Bendana2011}.

Sophisticated experimental setups have been built to investigate visible photons in the TEM. The photons can be analyzed spectrally, but also based on their polarization and emission angle \cite{takeuchiVisualizationSurfacePlasmon2011, coenenDirectionalEmissionPlasmonic2011, Bicket2018}. By employing Hanbury Brown-Twiss interferometry \cite{meuretPhotonBunchingReveals2017, tizeiSpatiallyResolvedQuantum2013, Scheucher2022}, it is possible to extract photon statistics and reveal properties of CL light. These advances benefit greatly from highly efficient single-photon detection. Recently, another significant leap forward has been made by introducing time-resolved direct electron detectors. The simultaneous single-photon and single-electron detection sensitivity and fast ($\sim\mathrm{ns}$) signal readout allow us to time tag individual electrons and thus identify coincident, correlated electron-photon pairs.
Previous works have exploited correlations within the electron-photon pairs to improve the signal-to-noise ratio in X-ray spectroscopy inside an electron microscope \cite{kruitDetectionXraysElectron1984,jannisSpectroscopicCoincidenceExperiments2019,jannisCoincidenceDetectionEELS2021}, increase the signal in excitation lifetime analysis \cite{yanagimotoTimecorrelatedElectronPhoton2023}, map photonic modes \cite{feistCavitymediatedElectronphotonPairs2022} and resolve the energy dependence of competing excitation pathways \cite{varkentinaCathodoluminescenceExcitationSpectroscopy2022} on the single electron-photon level.

In this Letter, we use coincidence detection of electrons and photons to investigate energy-momentum conservation within electron-photon pairs emerging in the ``coherent'' CL process. By imaging the electron recoil in coincident electron-photon pairs, we obtain a distribution that reflects the momentum distribution of the detected photons. We show that this result can be understood as a direct consequence of momentum conservation between the two particles. Although this fundamental property is firmly established in theory, it has so far eluded direct experimental observation. 
However, it is of crucial importance for the investigation of electron-photon pairs that share a common state in momentum-position space, going beyond classical correlations to explore entanglement \cite{einsteinCanQuantumMechanicalDescription1935,Konecna2022,henke2024electron-photon-entangl,kazakevich2024electron-photon-entangl}.

\begin{figure*}[hbt]
	\centering
    \includegraphics[width=\textwidth]{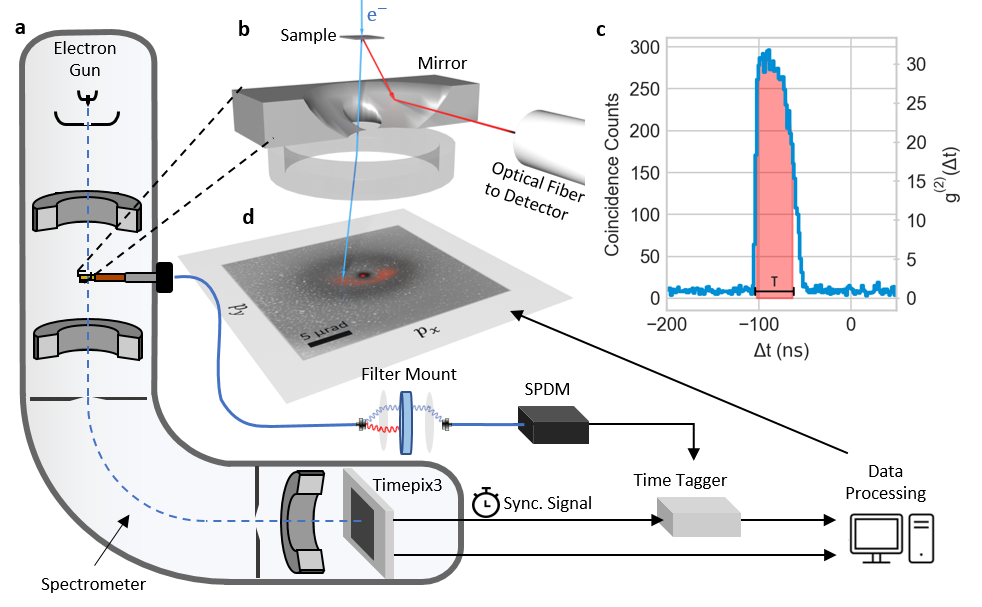}
	\caption{Experimental setup for detecting coincident electron-photon pairs. (a) A schematic of a TEM with an electron energy filter (spectrometer), where 200 keV electrons interact with a sample (thin silicon membrane). The electrons are detected in diffraction mode with a TP3-based camera and time-stamped. The corresponding CL photons are collected by a Gatan Vulcan sample holder (b) and guided through a multimode fiber to a single photon detection module (SPDM). A time tagger ensures the necessary synchronisation, which enables us to correlate individual photons to their originating electrons in time, clearly visible in a $g^{(2)}$ plot (c) at a certain time delay. These electron-photon pairs can now be analysed collectively, e.g. in terms of energy and momentum (d). 
}
	\label{fig:Setup}
\end{figure*} 

Fig.~\ref{fig:Setup}(a) gives a general overview of the experimental setup. In a TEM (FEI/TFS Tecnai G2 F20) a continuous beam of electrons with a primary beam energy of $E_\mathrm{kin}= 200\; \mathrm{keV}\pm0.45 \;\mathrm{eV}$ (FWHM) and a current of $\sim 3$~pA irradiates the sample. Each electron has a certain probability of coherently producing a CL photon at the interaction time $t = 0$. Independently, the electron can excite the sample to incoherently emit a photon after a random delay, depending on the characteristic lifetime of the process \cite{Meuret2015}. The Gatan Vulcan CL TEM holder features two ellipsoidal mirrors, one above and one below the sample, the bottom mirror is schematically depicted in Fig.~\ref{fig:Setup}(b). These mirrors are aligned to collect photons from a focal point at the sample plane and reflect them onto the faces of two separate multimode fibers, which guide the photons to a single-photon detection module (SPDM). Optionally, an optical filter can be applied to investigate only photons in a certain energy range. Each photon detection event is time stamped by the time tagging device, resulting in a recorded value $t_\gamma$. The electrons transmitted through the sample further propagate through the microscope. Its magnetic lenses are set to detect the electrons in the diffraction plane, as shown in Fig.~\ref{fig:Setup}(b,d). This configuration is known as diffraction mode. Optionally, the electrons can be energy-filtered using a spectrometer (Gatan GIF Tridiem, post-column imaging filter). Finally, they are directed to a TimePix3 (TP3) direct electron detection camera, which can detect the time of impact of individual electrons, resulting in a time stamp $t_\mathrm{e}$ for each of them. The arrival times $t_\gamma$ and $t_\mathrm{e}$ can be correlated in post-processing, resulting in a temporal cross-correlation histogram [an example shown in Fig.~\ref{fig:Setup}(c), see supplementary information for more detail]. For various reasons (e.g., TP3 detector jitter and/or synchronisation between electron and photon detection) our temporal resolution is limited to about 50~ns. To match electrons with coincident photons, we determine the difference in detection time between each photon ($t_\gamma$) and the electron ($t_\mathrm{e}$) that matches the expected time delay most closely. A pair of detection events is classified as a coincidence event if the time difference is within $\pm\tau/2=\pm 25\;\mathrm{ns}$ to the expected value $\mathbb{E}[\Delta t_\mathrm{e\gamma}]= \mathbb{E}[t_\mathrm{e}-t_\mathrm{\gamma}]$.

We let the electrons interact with a $100\;\mathrm{nm}$ thick monocrystalline silicon membrane ($500\times500\,\upmu \mathrm{m}^2$, in (100)-orientation, Silson Ltd.). Similar samples were shown to support a variety of electron-triggered photon production mechanisms, which were observed using both CL \cite{yamamotoCherenkovTransitionRadiation1991,yamamotoImagingCherenkovTransition1995,yamamotoCherenkovTransitionRadiation1997, stoger-pollachFundamentalsCathodoluminescenceSTEM2019} and momentum-resolved electron energy-loss spectroscopy (q-EELS) \cite{chenElectronenergyLossesSilicon1975,stoger-pollachCerenkovLossesLimit2006, saitoOpticalGuidedModes2013,erniImpactSurfaceRetardation2008}. We configure the electron microscope to illuminate the Si membrane with a nearly parallel electron beam (diameter $\sim50\,\upmu\mathrm{m}$ at the sample plane) to produce a low-angle diffraction (LAD) image of the zero-order Bragg peak. If the diffraction image is energy-filtered, it clearly exhibits a non-trivial pattern related to the distribution of electron momenta due to the excitation of photons, as demonstrated in the ring-like pattern shown in Fig.~\ref{fig:Setup}(d).

\begin{figure}[tb]
	\centering
    \includegraphics[width=0.5\textwidth]{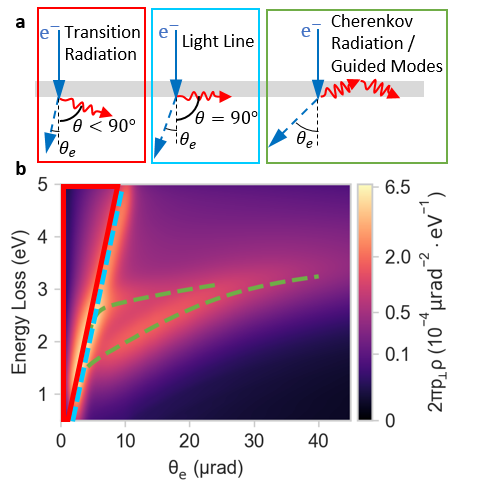}
	\caption{Visualization of the energy-loss and angular deflection probability distribution for electrons passing through the silicon membrane. The schematics (a) represent the various CL production mechanisms that contribute to the signal according to their characteristic dispersion relations and photon emission process (see the main text). (b) Displays the calculated probability density of specific energy losses corresponding to angular deflections $\theta_\mathrm{e}$ of electrons. The overlaid dashed lines trace the most probable paths for different processes.} 
	\label{fig:Prediction}
\end{figure} 

To interpret the experimental results, we first calculate the probability density $\rho(E,\mathbf{p}_\perp)$ for an electron that undergoes an energy loss $E$ while acquiring the transverse momentum $\mathbf{p}_\perp=(p_x,p_y)$. This can be used to infer the electron deflection angle $\theta_\mathrm{e}=p_\perp/p_z$, where $p_z$ is the momentum along the electron-beam axis. This calculation, shown in Fig.~\ref{fig:Prediction}(b), considers the coherent and weak electron-sample interaction. The model (see supplementary material) embeds the scattering geometry (i.e., electrons moving along a straight trajectory and passing through the 100 nm thin film) and optical properties of silicon characterized by the dielectric function \cite{schinke2015Sieps}. The energy and transverse momentum in the modeled coherent electron-photon scattering process are conserved: $E=\hbar\omega$, where $\hbar\omega$ is the photon energy, and $\mathbf{p}_\mathrm{\perp,f}-\mathbf{p}_\mathrm{\perp,i}=-\hbar\mathbf{k}_\perp$, with $\mathbf{p}_\mathrm{\perp,f/i}$ being the final/initial electron transverse momentum and $\mathbf{k_\perp}$ being the transverse component of the photon wave vector. Importantly, due to the energy-(transverse) momentum conservation, $\rho$ can be directly mapped on the distribution of the photons emerging in the electron-sample interaction.

The energy-momentum conservation also allows us to assign two distinct scenarios to separate regions of the probability density $\rho$. Cherenkov emission \cite{garciadeabajoBoundaryEffectsCherenkov2004,chenElectronenergyLossesSilicon1975, stoger-pollachCerenkovLossesLimit2006, erniImpactSurfaceRetardation2008,saitoOpticalGuidedModes2013} produces photons propagating inside the Si membrane and causes transverse momentum transfer to the electron below the ``light line'', which is defined by $p_\perp = \hbar k_0$, with $k_0=\omega/c$ and the speed of light in vacuum $c$, see Fig.~\ref{fig:Prediction}(a). In our sample, as a result of the large refractive index $n\approx 4$, Cherenkov photons couple exclusively to guided modes, which do not contribute to the radiative CL signal \cite{krogerBerechnungEnergieverlusteSchneller1968,garciadeabajoBoundaryEffectsCherenkov2004, saitoOpticalGuidedModes2013,yamamotoCherenkovTransitionRadiation1991}. On the other hand, the transition radiation consists of photons propagating outside the thin film, emitted at an angle $\theta$ with respect to the beam axis. Excitation of transition radiation requires a transverse momentum transfer $p_\perp = \sin{\theta}\,\hbar k_0$ from the electron to the photon. 
Consequently, transition radiation photons are always associated with electrons detected to the left of the light line, i.e., with transverse momentum $p_\perp<\hbar k_0$.

Before presenting our coincidence data, we need to understand the setup that constrains both electron and photon detection. As mentioned previously, we can employ a spectrometer to energy-filter the transmitted electrons. The resulting measured momentum-dependent probability density is then expressed as $\rho(E,\mathbf{p}_\perp)\alpha(E)$, where $\alpha(E)$ is a function that reflects the transmission and rejection of electrons depending on their energy loss. Detection of the far-field photons is slightly more complicated. Our CL setup utilizes ellipsoidal mirrors to collect and fiber-couple CL photons. The mirrors accept photons only from a certain solid angle, as visualized by the 3D plot of the mirror geometry in Fig.~\ref{fig:Setup}(b) (see supplementary material). 
We express this geometry limitation of photon collection as well as the efficiency differences in guiding and detecting photons of various momenta and energies with a function $\alpha_\gamma(\hbar\omega,\hat{\kb}_\perp)$, where $\hat{\kb}_\perp=(k_x/k_0,k_y/k_0)$.  

We first analyse the experimental data as a conventional LAD image, without coincidence filtering and impose energy-filtering on the electron side (see supplementary material for experimental details).
Fig.~\ref{fig:EFTEM}(a) shows data that is only minimally filterd (left), compared to narrowly energy-filtered LAD images. The energy filtering is performed within intervals with a width of $2\delta =1$~eV, each having a different central value $E_\mathrm{c}$, which imposes $\alpha=\mathrm{rect}[( E- E_\mathrm{c})/(2\delta)]$. The dark spot in the middle of the images represents the portion of the undeflected parallel probing beam that passes the energy filter due to the tail of the initial energy distribution of the electron source. The remaining part of the energy-filtered LAD images can be interpreted as $I(\pb_\perp)\propto\int dE\,\rho(E,\pb_\perp)\alpha(E)$, describing the aggregate distribution for all electrons with an energy-loss value within the accepted range. We observe a dominant ring-like feature, which we attribute to the excitation of CL photons, propagating nearly parallel to the slab's surface. The perimeters of the rings (denoted by red circles) are always located at the light line, i.e. $p_\perp\approx E_\mathrm{c}/c$, which confirms the theoretical prediction of maxima in the energy-momentum dependence of the loss probability in Fig.~\ref{fig:Prediction}(b). By filtering at higher energy losses, we see larger rings due to higher momentum transfer, allowing us to investigate the dispersion relation of the underlying CL mechanism. Since these data show rotational symmetry around the center determined by the undeflected part of the beam. We utilize integration over the azimuthal angle $\phi_\mathrm{e}$ to obtain the radial profile as $I(p_\perp)=p_\perp \int d\phi_\mathrm{e} I(\pb_\perp)$, where $\phi_\mathrm{e}=\mathrm{arctan}(p_y/p_x)$. The integration increases the signal-to-noise ratio (SNR) and facilitates comparison with the theory, see Fig.~\ref{fig:EFTEM}(c) and blue histograms/lines therein.
When accounting for the imperfect removal of the zero-loss beam, the data (blue shaded area) is in excellent agreement with the theory (solid blue line), considering contributions from all the coherent interaction mechanisms discussed above. This is what we can learn from energy-filtered electron images: we see a distribution of energy loss and deflection that results from a variety of superimposed loss mechanisms with no direct way of identifying which process contributes to the radiative CL signal.

\begin{figure*}
    \centering
	\includegraphics[width=\textwidth]{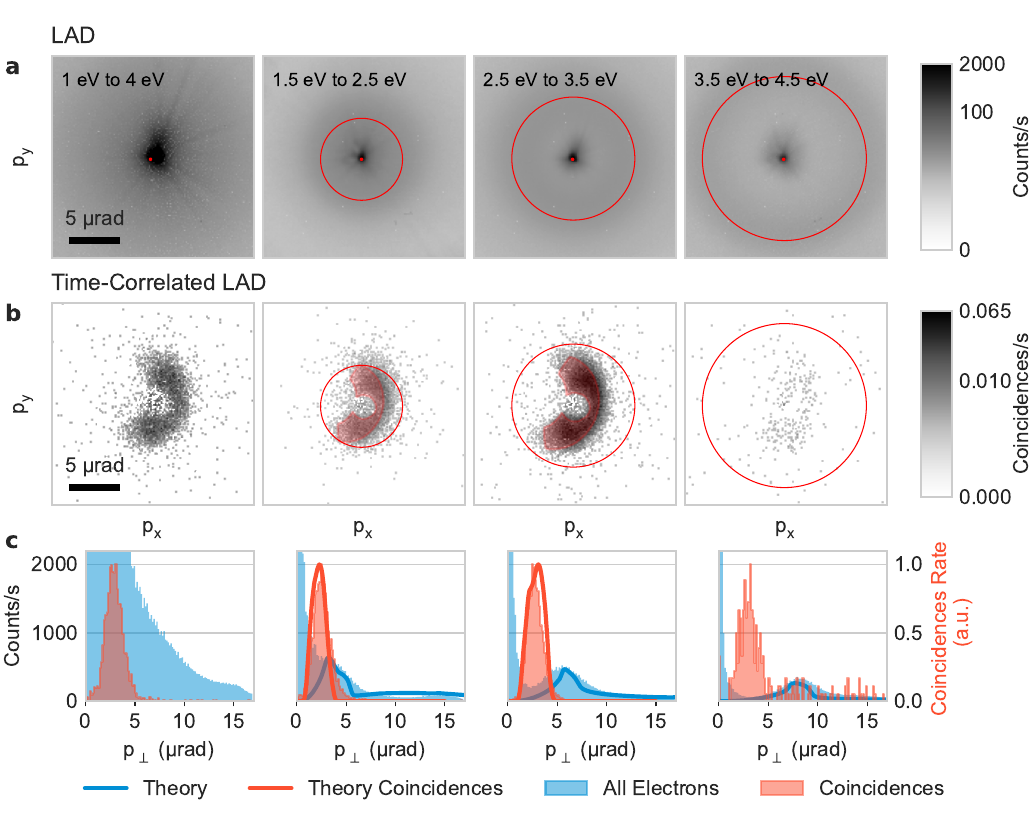}
	\caption{Energy-filtered low-angle diffraction images are shown without (a) and with (b) coincidence filtering applied. While the light line, as theoretically expected at the red ring, is the most prominent feature in the unfiltered distribution, additional structure corresponding to electron-photon pairs with energy-momentum inside the light line is revealed by applying coincidence filtering. In (c) the integrated radial profiles around the marked center are compared to the theoretical predictions.}
	\label{fig:EFTEM}
\end{figure*} 

In the following step, we match electrons to photons based on their arrival time and chart their counts as a function of the electron deflection angle, as shown in the red histograms in Fig.~\ref{fig:Setup}(c). A distinctive coincidence peak (FWHM = 42 ns) is visible, correlating mainly coherent far-field-emitted CL photons to electrons. A background due to uncorrelated coincidences is present. We subsequently extract the energy- and coincidence-filtered electron images in Fig.~\ref{fig:EFTEM}(b).  
This leads to a clearly discernible signal, which differs significantly from the raw images in  Fig.~\ref{fig:EFTEM}(a): the central spot, an artefact from the initial energy spread, is blanked out completely since these electrons did not produce a photon. 
The majority of electron counts show momenta, which are inside the red circles denoted in  Fig.~\ref{fig:EFTEM}(a). This is characteristic for electrons that produce far-field transition radiation, see Fig.~\ref{fig:Prediction}. On closer inspection, we see that although the sample and beam are rotationally symmetric about the beam axis, this symmetry is broken for the coincidence image. Instead, the characteristic horseshoe-like shape of the photon collection mirror is revealed, as highlighted by the light red area. This is initially surprising as the CL mirror is located in the photon path and does not influence the electrons directly. At a closer look, we can understand the distribution of the electron counts using the predicted transverse momentum conservation between electron and photon for this sample, resulting in $I_\mathrm{coin}(\pb_\perp)\propto\int dE\,\rho_\mathrm{TR}(E,\pb_\perp)\alpha( E)\alpha_\gamma(E,-\pb_\perp c/E)$. We defined $\rho_\mathrm{TR}$ as the probability density that an electron undergoes an energy loss $E$ and acquires the transverse momentum $\pb_\perp$ while also producing a photon which propagates to the far field in the forward direction with respect to the electron's propagation (unlike $\rho$, which comprises all coherent CL processes). Notably, contributions from incoherent CL are not included, this is justified, as we consider them small compared to the coherent contributions for this sample (see supplementary material for more detail).  

\begin{figure}[h!]
	\centering
    \includegraphics[width=0.65\textwidth]{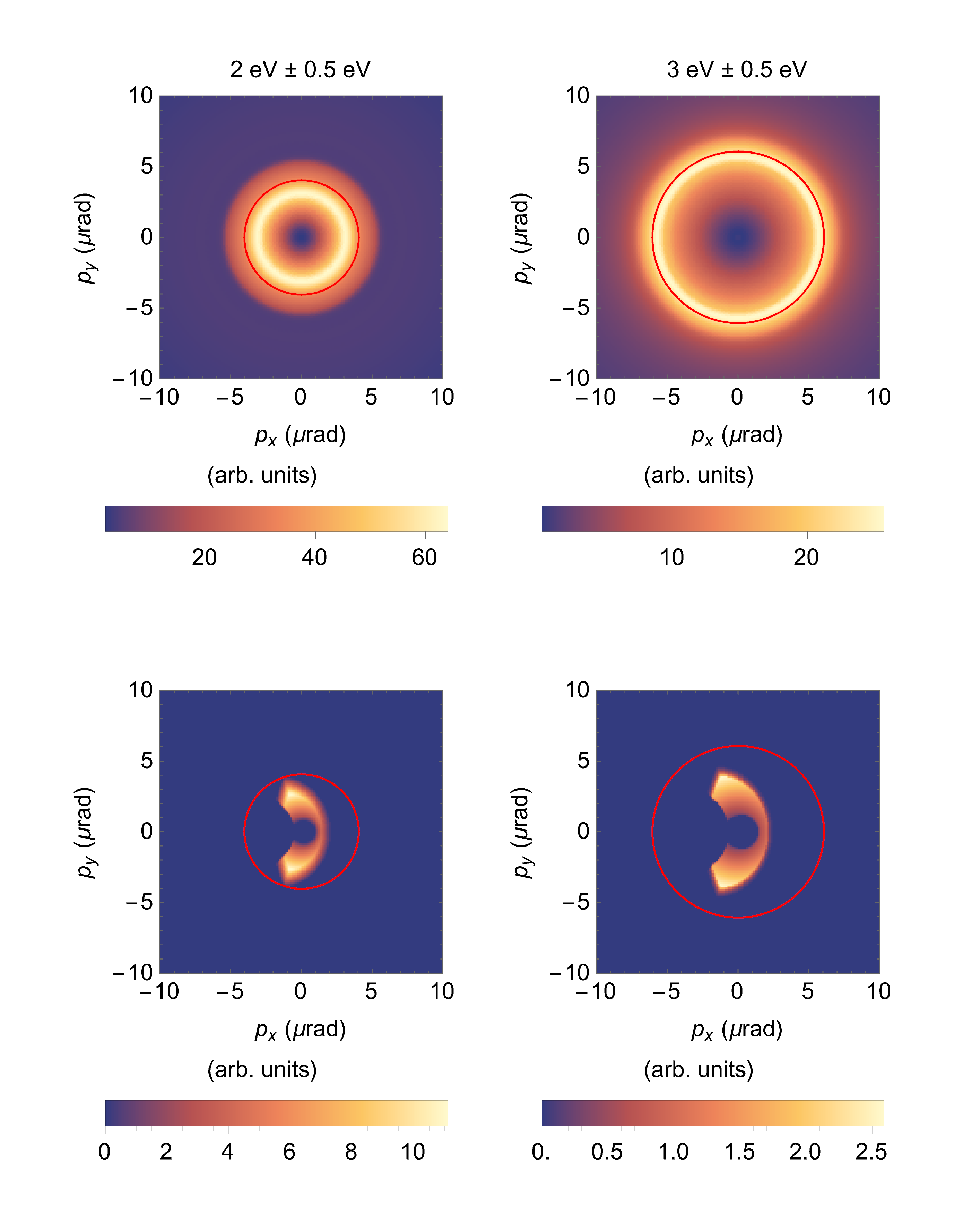}
	\caption{Simulated energy-filtered LAD images without (upper row) and with (lower row) coincidence filtering. For comparison with the experimental results in Fig.~\ref{fig:EFTEM}, we sum contributions in energy intervals corresponding to the energy-filtering windows with central energies 2 and 3~eV (left vs. right column). The red circles denote the boundary between radiative and non-radiative processes determined by the light line and at the central energy.}
	\label{fig:LAD_theory}
\end{figure}

Explained in terms of momentum conservation between electron and photon: if a photon is emitted to the azimuth angle $\phi$, the corresponding electron is deflected in the opposite direction $\phi + \pi$. The mirror does not cover all azimuth angles (see Fig.~\ref{fig:Setup}(b) and supplementary material), therefore electrons are unlikely to contribute to the coincidence image unless they produce photons accepted by the mirror. The coincidence-filtered LAD images thus have to trace the shape of the photon-collection region in momentum space. Further restrictions are imposed by the fiber and detector, which only collect photons in a restricted energy range (see supplementary material). 
This low detection efficiency in the UV range is likely the reason for the markedly decreased coincidence signal for $E_\mathrm{c}=4$~eV. 
In order to account for instabilities in coincidence matching we show normalised coincidence data in Fig.~\ref{fig:EFTEM}(c) and Fig.~\ref{fig:Photon_filtered}(c) (see supplementary information). 
Theoretical predictions are in good agreement with our data (see also Fig.~\ref{fig:LAD_theory}). 

\begin{figure*}
	\centering
	\includegraphics[width=\textwidth]{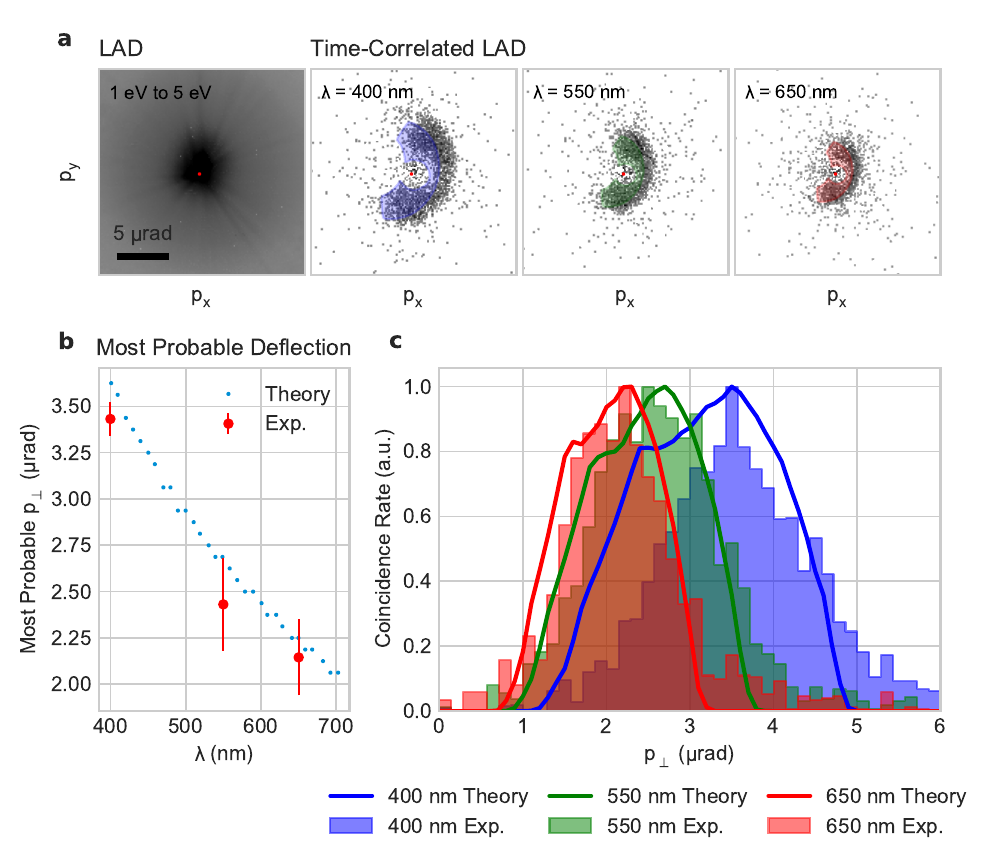}
	\caption{Coincidence-filtered low-angle diffraction images in (a) are produced by applying different bandpass wavelength filters ($\mathrm{FWHM} =40$~nm @ 400 nm, 550 nm, 650 nm) to the photons. The radial profiles (c) are obtained by integrating the data over the azimuthal angle $\phi$. They include theoretical predictions, which agree well with the dispersion relation of transition radiation. In (b) the most probable deflection value for each bandpass filter is compared to the value predicted by the theoretical model. The error bars provide an estimate of the statistical uncertainty.}
	\label{fig:Photon_filtered}
\end{figure*} 

Rather than applying an energy filter on the electron side, one can use a wavelength filter on the photon side. The measurements shown in Fig.~\ref{fig:Photon_filtered} were performed using band-pass filters with a full width at half maximum of $2\delta_\lambda=40\;\mathrm{nm}$ at different central wavelengths $\lambda_\mathrm{c}$. This results in a modified function $\alpha'_\gamma(\hbar\omega,\hat{\kb}_\perp)=\alpha_\gamma(\hbar\omega,\hat{\kb}_\perp)\mathrm{rect}[(2\pi/k_0-\lambda_\mathrm{c})/\delta_\lambda]$. The coincidence LAD images in Fig.~\ref{fig:Photon_filtered}(a) show the same horseshoe-like pattern as before and scale with the photon momentum. The wavelength distribution still reflects the energy-momentum conservation between the electron-photon pairs and can be understood with the model distribution $I_\mathrm{coin}(\pb_\perp)=\int dE\,\rho_\mathrm{TR}(E,\pb_\perp)\alpha'_\gamma(E,-\mathbf{p}_\perp c/E)$. The theoretically predicted coincidence radial profiles [color-coded lines in Fig.~\ref{fig:Photon_filtered}(c)], as well as the most probable electron deflection as a function of $\lambda_\mathrm{c}$ in Fig.~\ref{fig:Photon_filtered}(b), show a qualitative agreement with the experimentally obtained data.

We have presented energy- and momentum-resolved coincidence measurements of electron-photon pairs, along with a theoretical description that naturally follows from the assumption of energy and momentum conservation. According to this theory, it should be possible to directly map the angular dependence of photon detection efficiency, as determined by the shape of the collecting mirror, onto the momentum distribution of the coincident electron.
We are the first to demonstrate this effect -- known in the photonics community as ``ghost imaging'' \cite{pittmanOpticalImagingMeans1995,padgettIntroductionGhostImaging2017, rotunnoOneDimensionalGhostImaging2023} -- using electron-photon pairs, thereby also confirming the underlying phenomenon of energy-momentum conservation in coherent CL processes.

While previous momentum-resolved EELS studies showed all energy-loss mechanisms and their respective momentum transfer superimposed, combined energy- and coincidence filtering allowed us to clearly separate processes which contribute to the CL signal, and therefore must have resulted in the emission of a far-field photon. For electrons emitting transition radiation, we saw an overall enhancement by a factor of $\approx 17$ compared to the signal without coincidence-filtering (i.e., energy filtering only, see supplementary material). This produced a clearly discernible signal which had been completely hidden in previous EELS studies. The improvement is especially pronounced in the presence of strong background signals that don't contribute to the CL. An example of this is the low-energy tail of the electron source that was strongly visible in the energy-filtered image in Fig.~\ref{fig:EFTEM}(a) but completely suppressed after coincidence filtering in Fig.~\ref{fig:EFTEM}(b). 

Moreover, we argued that information about the correlated electron can be used to enhance the capabilities of CL spectroscopy. We can trace the origin of CL-photons back to their production mechanism by comparing the momentum transfer on the electron to the characteristic dispersion relation of the photon. 

We also highlight that the presented technique combines the strengths of q-EELS and CL spectroscopy when looking at radiative coherent processes. The use of optical spectroscopy generally allows for a higher energy resolution than electron energy-loss spectroscopy. As the energy loss can be measured directly on the photon instead of measuring the final energy of the electron, the resulting distribution is independent of the initial energy spread of the beam. 
Simultaneously, we obtain precise knowledge of the corresponding electron momentum by using the diffraction mode in the electron microscope. These momentum correlations are particularly important for the study of entanglement in momentum-position space and might enable revolutionary techniques adopted from photonic quantum optics \cite{einsteinCanQuantumMechanicalDescription1935,Zou1991, Lemos2014ZWM_undetected_photons} for modern electron microscopy.

\section*{Funding Information and Acknowledgments}
The authors thank Isobel Bicket, Michael Stöger-Pollach \& Santiago Beltrán-Romero for fruitful discussions. PH, AP thank the Austrian Science Fund (FWF): Y1121, P36041, P35953. This project was supported by the ESQ-Discovery Program 2020 "A source for correlated electron-photon pairs" and the FFG-project AQUTEM. AK and MH acknowledge the support of the Czech Science Foundation GACR under the Junior Star grant No. 23-05119M.

\bibliography{bib}

\onecolumngrid

\section{\textsc{Supplemental Material}}
\subsection{CL collection efficiency}
\begin{figure}[tb]
	\centering
    \includegraphics[width=0.8\textwidth]{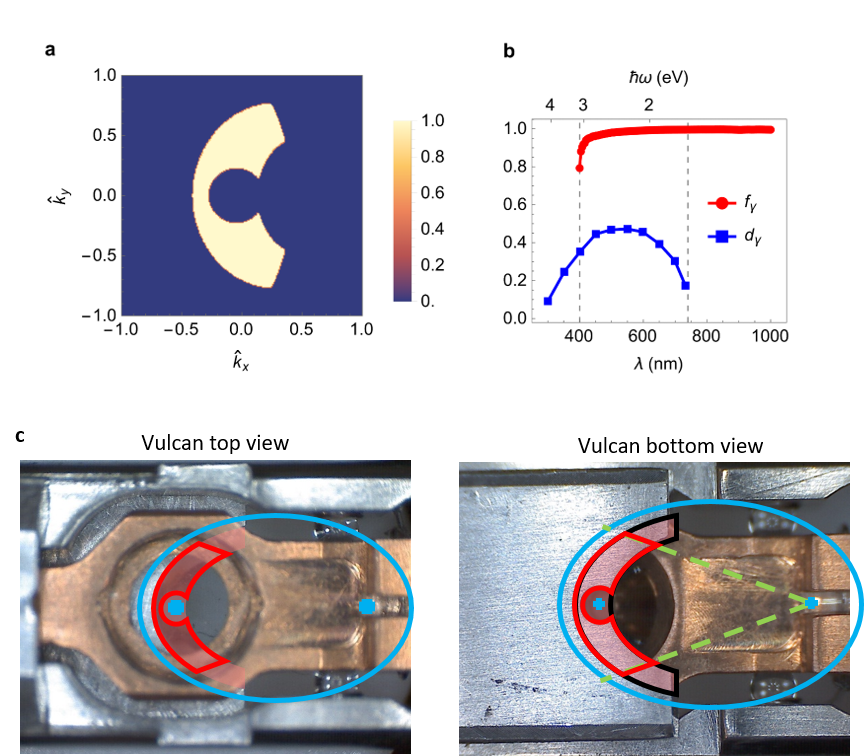}
	\caption{Restrictions on the photon collection imposed by the CL mirror, fiber, and detector. (a) Photon collection probability imposed by the elliptical mirror, (b) energy-dependent photon collection of the fiber (red line) and detector (blue line). (c) Images of the Gatan Vulcan CL sample holder, highlighting the reflective surface of the bottom mirror (red shaded area). The ellipsoid of revolution of the ellipse shown in blue with respect to the holder axis is used to model the shape of the mirror surface. Its focal points are marked with blue crosses.}
	\label{fig:Photon_collection}
\end{figure}

The setup's collected solid angle is mostly determined by the shape of a specially designed ellipsoidal mirror schematically shown in Fig.~\ref{fig:Setup}(b) and in detail in Fig.~\ref{fig:Photon_collection}(c). The mirror is milled from a piece of aluminum that has plane, parallel surfaces at its top and bottom. It extends approximately $400\;\mathrm{\upmu m}$ in the z-direction, forming a section of an ellipsoid that is too shallow to be closed around the beam hole. In the top view, this results in a horseshoe-shaped mirror surface. Parts of the mirror that are far off the holder axis are shaded by the holder, as indicated in Fig.~\ref{fig:Photon_collection}(c). Consequently, the effective mirror area is reduced. 

As the illuminated area on the sample is small ($\approx 50\;\mathrm{\upmu m}$ diameter) compared to the size of the mirror (focal length $\approx2\;\mathrm{mm}$), it can be regarded as a point source in the mirror's focal point. We can model the photon collection efficiency $A_\gamma(\hbar\omega,\hat{\kb}_\perp)$ considering the mirror geometry, extracted from a series of measurements on the holder, as well as the finite numerical aperture of the fiber. The obtained function $A_\gamma(\hat{\kb}_\perp)$ plotted in Fig.~\ref{fig:Photon_collection}(a) shows a nearly homogeneous photon collection probability in a range of normalized transverse momenta (scattering directions) restricted by the mirror edges. Further restrictions arise due to the wavelength-dependent efficiency of the guiding fiber and detector, as expressed by functions $f_\gamma(\hbar\omega)$ and $d_\gamma(\hbar\omega)$ plotted in Fig.~\ref{fig:Photon_collection}(b). The overall photon detection efficiency can be expressed as $\alpha_\gamma(\hbar\omega,\hat{\kb}_\perp)=A_\gamma(\hat{\kb}_\perp)f_\gamma(\hbar\omega)d_\gamma(\hbar\omega)$.

\subsection{Joint electron-energy-loss and photon-detection probability}
Previous works \cite{jannisSpectroscopicCoincidenceExperiments2019,feistCavitymediatedElectronphotonPairs2022} have modeled the coincidence count rate simply as $R_\mathrm{coin}= i\alpha_\mathrm{e}\alpha_\mathrm{\gamma}p$ with the current $i$ in electrons/second,  $\alpha_\mathrm{e}$ being the probability to detect the electron (limited by detector efficiency and scattering out of our limited collection angle), $\alpha_\gamma$ being the photon detection probability. The photons can be produced coherently or incoherently with probabilities $p_\mathrm{coh}$ and $p_\mathrm{in}$, and both types of mechanisms contribute to the overall photon production probability $p=p_\mathrm{coh}+p_\mathrm{in}$. 

In order to extend this idea, we need to consider the energy and angular dependence of the photon production mechanisms and the collection efficiencies:
\begin{align*}
\alpha&\xrightarrow{}\alpha(E,\mathbf{p}_\perp),\\
\alpha_\mathrm{\gamma}&\xrightarrow{}\alpha_\mathrm{\gamma}(\hbar\omega,\hat{\kb}_\perp),\\
p&\xrightarrow{} \rho(E, \pb_\perp,\hbar\omega,\hat{\kb}_\perp)
\end{align*}
with $\rho$ being a probability density of detecting an electron with energy loss $E$ and momentum transfer $\pb_\perp$, and a photon of energy $\hbar\omega$ and normalized transverse momentum $\hat{\kb}$. The probability density can be split into coherent and incoherent contributions, however, we neglect the latter and assume that we only collect photons produced by the coherent process, i.e. $\rho\rightarrow\rho_\mathrm{coh}$. We further consider the energy and transverse momentum conservation in the electron-photon pairs, which we express in terms of conditional probability 
\begin{align}
\rho_\mathrm{coh}(E,\mathbf{p}_\perp|\hbar\omega,\hat{\kb}_\perp)= \rho_\mathrm{coh}(E,\mathbf{p}_\perp,\hbar\omega,\hat{\kb}_\perp)\delta^2(\mathbf{p}_\perp+\hbar k_0 \hat{\kb}_\perp) \delta(E-\hbar\omega). 
\end{align}

In our case, the coherently emitted photons correspond to the transition radiation. We thus express the probability density of the transition radiation $\rho_\mathrm{TR}$ formally as an integral of $\rho_\mathrm{coh}$ over the collected photon energies and momenta
\begin{align}
    \rho_\mathrm{TR}(E,\pb_\perp)=\int d(\hbar\omega)\, d\hat{\kb}_\perp\,\rho_\mathrm{coh}(E,\mathbf{p}_\perp,\hbar\omega,\hat{\kb}_\perp)\delta^2(\mathbf{p}_\perp+\hbar k_0 \hat{\kb}_\perp) \delta(E-\hbar\omega).
\end{align}
To simulate the experimentally detected coincidences, we need to consider the energy- and momentum-dependent collection probability on the photon side $\alpha_\gamma(\hbar\omega,\hat{\kb}_\perp)$ as well as the possibility of energy filtering on the electron side $\alpha(E)$:
\begin{align}
    \rho_\mathrm{exp}(E,\pb_\perp)&=\alpha(E)\int d(\hbar\omega)\, d\hat{\kb}_\perp\,\alpha_\gamma(\hbar\omega,\hat{\kb}_\perp)\rho_\mathrm{coh}(E,\mathbf{p}_\perp,\hbar\omega,\hat{\kb}_\perp)\delta^2(\mathbf{p}_\perp-\hbar k_0 \hat{\kb}_\perp) \delta(E-\hbar\omega)\nonumber\\
    &=\rho_\mathrm{TR}(E,\pb_\perp)\alpha(E)\alpha_\gamma(E,-\pb_\perp/(\hbar k_0))\nonumber\\
    &=\rho_\mathrm{TR}(E,\pb_\perp)\alpha(E)f_\gamma(E)d_\gamma(E)A_\gamma(-\pb_\perp c/E).
    \label{eq:exp_coincidence_rho}
\end{align}
The measured LAD intensities can then be modeled and interpreted according to the expressions in the main text, by evaluating integrals of Eq.~\eqref{eq:exp_coincidence_rho} over $E$. It is worth noticing that when the energy-filtering window does not coincide with energy regions where photons are efficiently collected [see Fig.~\ref{fig:Photon_collection}(b)], we should not observe any coincidence counts related to the coherent process. Moreover, we should observe the outline of the mirror in the electron momentum space, determined by the function $A_\gamma$ [plotted in Fig.~\ref{fig:Photon_collection}(a)].

\subsection{Theoretical description of (energy-filtered) low-angle diffraction images}

\subsubsection{Electric field in the interaction of fast electrons with a thin film}
To evaluate the probability of electron energy loss and far-field photon emission, we utilize the formalism based on classical electrodynamics \cite{Lucas1970}. We calculate the electromagnetic field excited due to point-like electrons traversing the thin film composed of an isotropic medium characterized by a dielectric function. 

In the following, we assume that energy exchange and the momentum transfer during the interaction are small compared to the electron central kinetic energy and momentum. We thus consider electrons moving at velocity $v$ along the positive $z$ axis, which generate the free current density $\Jb_\mathrm{F}=\rho_\mathrm{F}v\hat{z}$, where $\rho_\mathrm{F}=-e\delta(x)\delta(y)\mathrm{exp}(\ii\omega z/v)/v$ is the free charge density expressed in the frequency domain. The free charge and current density sources then enter the wave equation for the electric field $\Eb$, which is in the frequency domain and in isotropic, non-magnetic medium expressed as
\begin{align}
\left(\nabla^2+k_0^2\varepsilon\right)\Eb=\frac{1}{\varepsilon_0\varepsilon}\nabla \rho_\mathrm{F}-\ii\omega\upmu_0\Jb_\mathrm{F},
\label{Eq:wave_equation}
\end{align}
where $c=1/\sqrt{\varepsilon_0\upmu_0}$ is the speed of light in vacuum, $\varepsilon_0$ and $\upmu_0$ are the vacuum permittivity and permeability, respectively. The wave equation has to be solved separately inside the thin film (for $\lvert z\rvert \leq d/2$) characterized by the dielectric function $\epsilon(\omega)\neq 1$, and outside the thin film (for $\lvert z\rvert>d/2$). Due to the axial symmetry of the problem, it is beneficial to Fourier-transform the wave equation \eqref{Eq:wave_equation} and rewrite it as
\begin{align}
\left(-\alpha^2+\frac{\partial^2}{\partial z^2}\right)E_R(Q,z,\omega)&=-\frac{e\ii Q}{\varepsilon_0\varepsilon v} \ee^{\ii\omega z/v}\label{Eq:wave_equation_ER}\\
\left(-\alpha^2+\frac{\partial^2}{\partial z^2}\right)E_z(Q,z,\omega)&=-e\,\ii\omega\ee^{\ii\omega z/v}\left(\frac{1}{\varepsilon_0\varepsilon v^2}-\upmu_0 \right),\label{Eq:wave_equation_Ez}
\end{align}
where $Q=\sqrt{q_x^2+q_y^2}$ is the transverse reciprocal coordinate, $E_R=\Eb\cdot\Qb/Q$, and $\alpha^2=Q^2-\varepsilon\omega^2/c^2$. The solutions of Eqs.~\eqref{Eq:wave_equation_ER} and \eqref{Eq:wave_equation_Ez} are straightforward. Together with Gauss's law dictating $\ii Q E_R+\partial E_z/\partial z =\rho_\mathrm{F}/(\varepsilon_0\varepsilon)$, we find that inside the dielectric film, the field components are
\begin{equation}
\begin{aligned}
E_{R2}(Q,\lvert z\rvert< d/2)&=A_2\ee^{\alpha z}+B_2\ee^{-\alpha z}+\frac{\ii Q e}{v\varepsilon_0\varepsilon\left(\omega^2/v^2+Q^2-\varepsilon\omega^2/c^2\right)}\ee^{\ii\omega z/v},\\
E_{z2}(Q,\lvert z\rvert< d/2)&=\frac{-\ii Q}{\alpha}\left(A_{2}\ee^{\alpha z}-B_2\ee^{-\alpha z}\right)+\frac{\ii \omega e\left(1-\varepsilon v^2/c^2\right)}{v^2\varepsilon_0\varepsilon \left(\omega^2/v^2+Q^2-\varepsilon\omega^2/c^2\right)}\ee^{\ii\omega z/v},
\end{aligned}    
\end{equation}
where $A_2$ and $B_2$ are constants. The electric field components in vacuum outside the film are
\begin{equation}
\begin{aligned}
E_{R1}(Q,z> d/2)&=B_1\ee^{-\alpha_0 z}+\frac{\ii Q e}{v\varepsilon_0\left(\omega^2/v^2+Q^2-\omega^2/c^2\right)}\ee^{\ii\omega z/v},\\
E_{z1}(Q,z> d/2)&=\frac{\ii Q}{\alpha_0}B_1\ee^{-\alpha_0 z}+\frac{\ii \omega e\left(1- v^2/c^2\right)}{v^2\varepsilon_0\left(\omega^2/v^2+Q^2-\omega^2/c^2\right)}\ee^{\ii\omega z/v},
\label{Eq:field_out}
\end{aligned}
\end{equation}
and
\begin{equation}
\begin{aligned}
E_{R3}(Q,z< -d/2)&=A_3\ee^{\alpha_0 z}+\frac{\ii Q e}{v\varepsilon_0\left(\omega^2/v^2+Q^2-\omega^2/c^2\right)}\ee^{\ii\omega z/v},\\
E_{z3}(Q,z< -d/2)&=\frac{-\ii Q}{\alpha_0}A_{3}\ee^{\alpha_0 z}+\frac{\ii \omega e\left(1- v^2/c^2\right)}{v^2\varepsilon_0 \left(\omega^2/v^2+Q^2-\omega^2/c^2\right)}\ee^{\ii\omega z/v},
\end{aligned}
\label{Eq:field_out3}
\end{equation}
where we defined $\alpha_0^2=Q^2-\omega^2/c^2$. Note that $\alpha_0^2>0$ in the sub-luminal region, however, it can also be purely imaginary for $Q<\omega/c$, which corresponds to the excitation of propagating radiative modes above the light line. The sign of $\alpha_0$ is then chosen such that the waves propagate away from the slab. The four constants $B_1, A_2, B_2, A_3$ can be determined from the boundary conditions at $z=\pm d/2$, where we require $E_R$ and $D_z$ to be continuous. Hence, the boundary conditions become
\begin{equation}
    \begin{aligned}
E_{R1}(Q,d/2)=E_{R2}(Q,d/2),& \qquad E_{z1}(Q,d/2)=\varepsilon E_{z2}(Q,d/2),\\
E_{R3}(Q,-d/2)=E_{R2}(Q,-d/2),& \qquad E_{z3}(Q,-d/2)=\varepsilon E_{z2}(Q,-d/2).
\end{aligned}
\label{Eq:Boundary_conditions}
\end{equation}

\subsubsection{Electron energy-loss probability}
\begin{figure}[tb]
	\centering
    \includegraphics[width=0.95\textwidth]{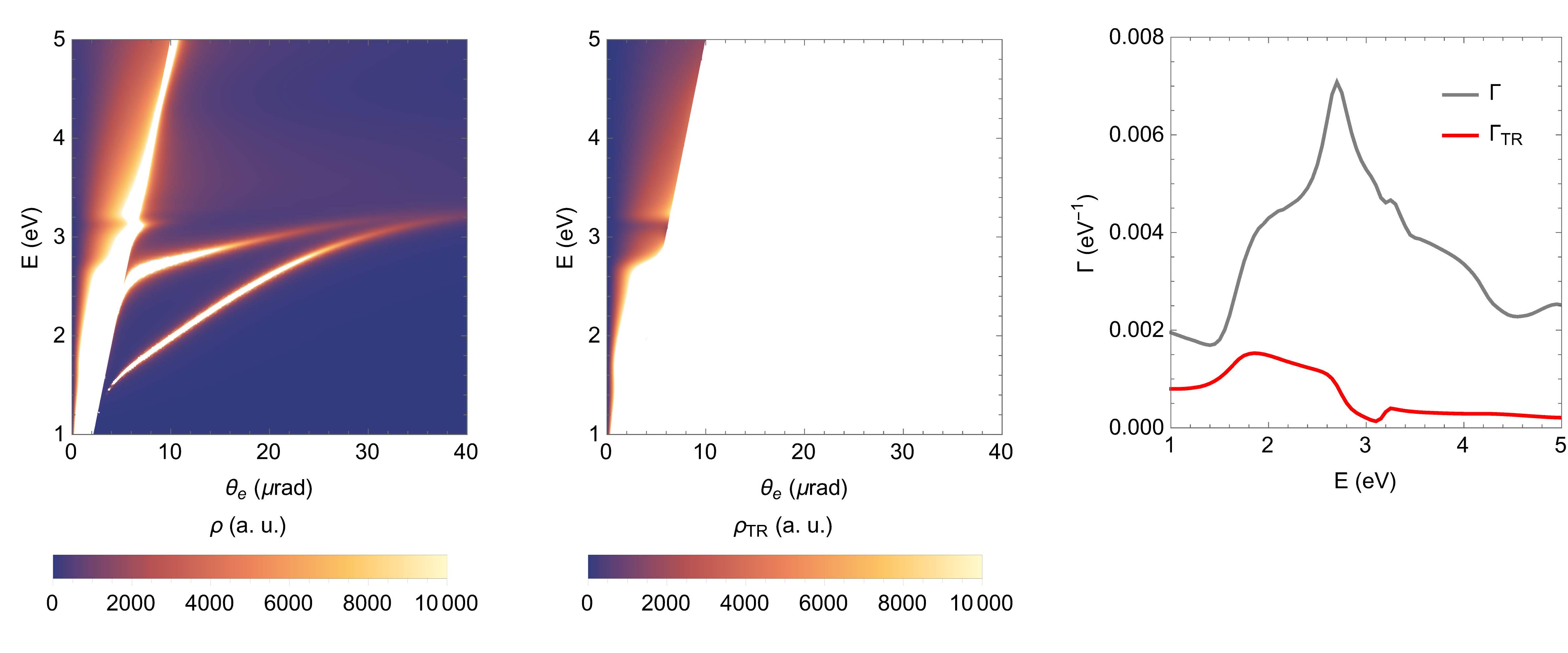}
	\caption{Calculated probability densities $\rho(Q,\omega)$ (left) and $\rho_\mathrm{TR}(Q,\omega)$ for photons propagating in the forward direction (center). Integration of the probability densities over the transverse momentum yields the electron energy-loss (gray) and transition radiation (red) probabilities compared in the right graph.}
	\label{fig:EELS_CL_theory}
\end{figure}

Electron energy-loss probability can be evaluated based on the classical energy loss due to the interaction with the induced electric field. As the external field produced by the electron in free space does not perform any work, the energy loss can be written in terms of the total field as
\begin{align}
\Delta E&=e\int_{-\infty}^\infty\mathrm{d}z\,E_z(\rb_\mathrm{e})=\mathrm{Re}\left\lbrace\frac{e}{4\pi^3}\int\limits_0^{\infty}\mathrm{d}\omega\ \int\limits_0^{\infty}\mathrm{d}Q\ 2\pi Q \int\limits_{-\infty}^\infty \mathrm{d}z\ E_z(Q,\omega,z)\ee^{-\ii\omega z/v}\right\rbrace.\label{Eq:DeltaE}
\end{align}
We now define the loss probability $\Gamma(\omega)$ as \cite{garciadeabajoOpticalExcitationsElectron2010a}
\begin{align}
\Delta E=\int_0^\infty\mathrm{d}\omega\,\hbar\omega\,\Gamma(\omega)=\int_0^\infty\mathrm{d}\omega\,\hbar\omega\,\int_0^\infty\mathrm{d}Q\,2\pi Q\,\rho(\omega,Q),
\label{Eq:rho_def}
\end{align}
where we recognize the probability density $\rho$ defined in the main text. Due to the symmetry of the problem, we can notice that $\rho$ is only a function of the magnitude of the transverse momentum transfer $p_\perp=\hbar Q$ and not its direction. By comparing right hand sides of Eqs.~\eqref{Eq:DeltaE} and \eqref{Eq:rho_def}, we obtain
\begin{align}
\rho(\omega,Q)&=\frac{e}{4\pi^3\hbar\omega}\mathrm{Re}\left\lbrace\int\limits_{-\infty}^{\infty}\mathrm{d}z\ E_z(Q,\omega,z)\ee^{-\ii\omega z/v} \right\rbrace\\
&=\frac{eQ}{4\pi^3\hbar\omega}\mathrm{Im}\left\lbrace\frac{A_{3}\ee^{-d/2(\alpha_0 -\ii\omega/v)}}{\alpha_0(\alpha_0 -\ii\omega/v)}-\frac{B_1 \ee^{-d/2(\alpha_0+\ii\omega/v) }}{\alpha_0(\alpha_0+\ii\omega/v)}+\frac{2 A_2\, \mathrm{sinh}\left[\frac{d(\alpha-\ii\omega/v)}{2}\right]}{\alpha(\alpha-\ii\omega/v)}-\frac{2B_2\,\mathrm{sinh}\left[\frac{d(\alpha+\ii\omega/v)}{2}\right]}{\alpha(\alpha+\ii\omega/v)}\right\rbrace\nonumber\\
&-\frac{d e^2}{4\pi^3\hbar v^2\varepsilon_0}\mathrm{Im}\left\lbrace\frac{1-\varepsilon v^2/c^2}{\varepsilon \left(\omega^2/v^2+\alpha^2\right)}\right\rbrace,\label{Eq:P(Q,om)}
\end{align}
where we can recognize that the last term corresponds to bulk losses in an unbounded medium while the preceding terms emerge due to the film boundaries. Eq.~\eqref{Eq:P(Q,om)} was used to generate the energy- and momentum-dependent electron energy-loss probability in Fig.~\ref{fig:EELS_CL_theory} (left plot), which after convolution in energy and momentum, taking into account the experimental resolution, yields Fig.~\ref{fig:Prediction}(b). We set the electron speed $v=0.695c$ according to the experimentally utilized acceleration voltage 200~kV, film thickness $d=100$~nm and the dielectric response of silicon based on data from Ref.~\cite{schinke2015Sieps}. We also note that $p_\perp=\hbar Q=q \,\mathrm{sin}\theta_\ee$, where $m_\ee$ is the electron rest mass, $q=m_\ee\gamma v$ is the electron wave vector and $\gamma=1/\sqrt{1-v^2/c^2}$ is the Lorentz contraction factor.

\subsubsection{Far-field photon emission (cathodoluminescence) probability}
As discussed in the main text and visualized in Fig.~\ref{fig:Prediction}(a), the electron energy-loss probability in Eq.~\eqref{Eq:P(Q,om)} contains different channels corresponding to photons not necessarily propagating to the far field. To express the electric field radiated from the thin film to the far field, we only consider the induced electric field $\Eb^\mathrm{ind}_{1}$ emitted in the $ z>d/2$ half-space, which can be evaluated by subtracting the non-radiative contribution of the external field of the fast electron in Eqs.~\eqref{Eq:field_out}. We express the induced field as a function of the real-space coordinates, which requires the Hankel transformation. Furthermore, taking the far-field approximation, we obtain \cite{Brenny2016_TR}

\begin{align}
\mathbf{E}_{1}^\mathrm{ind}(R,z>d/2)&=\frac{1}{2\pi}\int_0^\infty Q\mathrm{d}Q\,B_1\,\ee^{\ii q_z z}\left[\ii J_1(Q R)\RR-J_0(Q R)\,\frac{Q}{q_z}\zz \right]\nonumber\\
&\approx \frac{-\ii}{2\pi} \frac{\ee^{\ii k r}}{r}\,B_1\,\left[q_z\RR-Q\zz \right]=\frac{\ee^{\ii k r}}{r}\,\frac{-\ii k}{2\pi}B_1 \hat{\mathbf{\theta}}=\frac{\ee^{\ii k r}}{r}\,\mathbf{f}_1(\omega,\theta),
\label{Eq:E1indFF}
\end{align}
where we utilized $q_z=k\,\mathrm{cos}\theta$, $Q=k\,\mathrm{sin}\theta$ and $\mathrm{cos}\theta\,\RR-\mathrm{sin}\,\theta\zz =\hat{\theta}$, where $\hat{\theta}$ is the unit vector in spherical vector basis, and where the coefficient $B_1$ was evaluated using the boundary conditions [Eq.~\eqref{Eq:Boundary_conditions}]. We also defined the scattering vector $\mathbf{f}_1$, which depends only on the polar angle $\theta$ due to the symmetry. 

The cathodoluminescence probability, or in other words, the probability of the transition radiation emission of a photon with energy $\hbar\omega$ in the forward direction can be evaluated as 
\begin{align}
\Gamma_{\mathrm{TR}}(\omega)&=\frac{\varepsilon_0}{\pi\hbar k}\int_{\delta\Omega}d\Omega\,\lvert\mathbf{f}_1(\omega,\theta)\rvert^2\label{Eq:GamTR}.
\end{align}
The integration spans over a range of solid angles $\delta\Omega$ covered by the detector. We can now define the function $\rho_\mathrm{TR}(\omega,Q)$ as $\Gamma_{\mathrm{TR}}(\omega)=\int dQ\,2\pi Q\,\rho_\mathrm{TR}(\omega,Q)$, where
\begin{align}
    \rho_\mathrm{TR}(\omega,Q)=\frac{\varepsilon_0}{\pi\hbar k}\frac{1}{k q_z}\lvert\mathbf{f}_1(\omega,\mathrm{arcsin}(Q/k))\rvert^2=\frac{\varepsilon_0}{4\pi^3\hbar}\frac{1}{q_z}\lvert B_1(\omega,Q)\rvert^2.
\end{align}
Similarly, it is possible to define angle-dependent probability density as $\Gamma_{\mathrm{TR}}(\omega)=\int d\theta\,\mathrm{sin}\theta\,\rho'_\mathrm{TR}(\omega,\theta)$. The relation between momentum- and angle-dependent probability densities is then $ \rho'_\mathrm{TR}(\omega,\theta)=\rho_\mathrm{TR}(\omega, k\,\mathrm{sin}\theta)\,\mathrm{cos}\theta/(2\pi)$.

\subsection{Simulated LAD images}
\FloatBarrier

\begin{figure}[h!]
	\centering
    \includegraphics[width=0.9\textwidth]{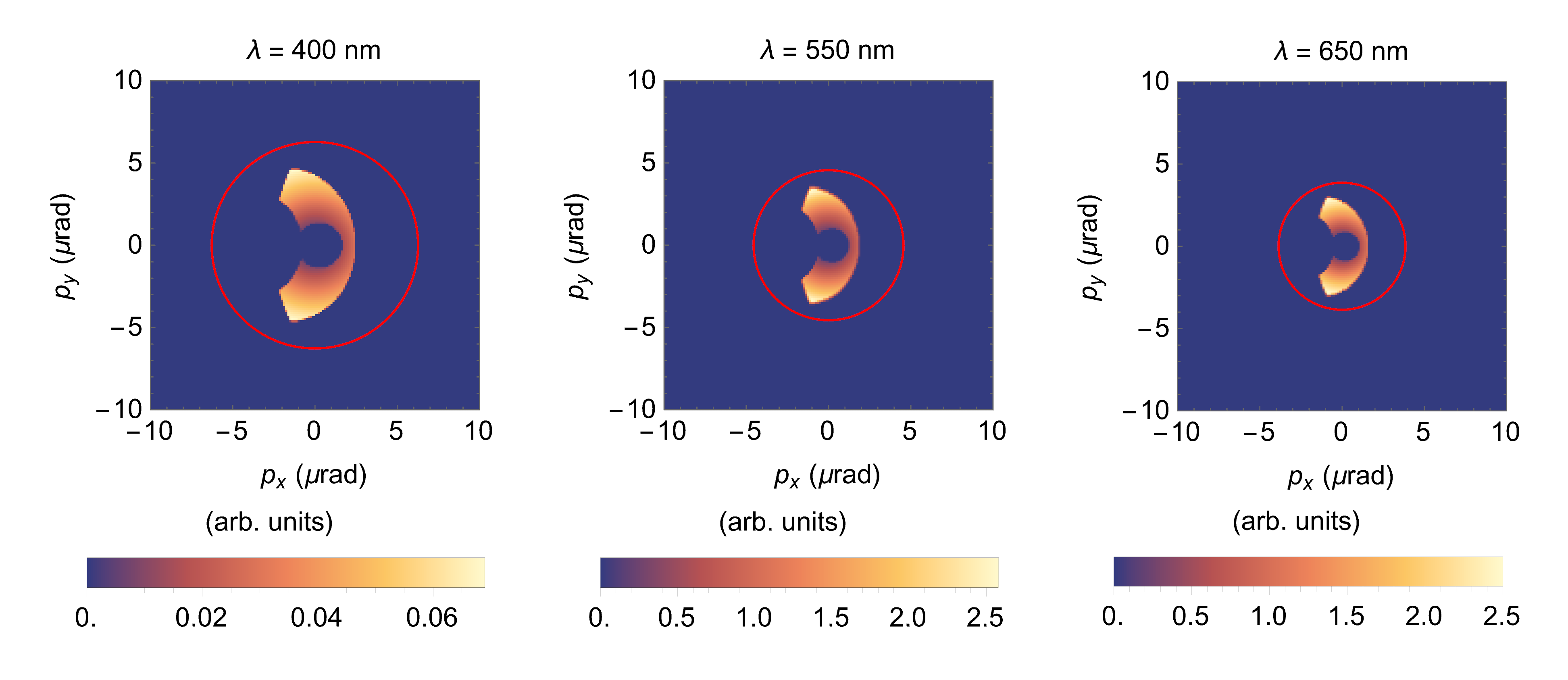}
	\caption{Simulated coincidence-filtered LAD images corresponding to the conditions in Fig.~\ref{fig:Photon_filtered}.}
	\label{fig:LAD_theory_photon}
\end{figure}

\FloatBarrier

\subsection{Contributions from Incoherent CL}\label{sec:incoherent_radiation_negligible}
According to the measurements presented in this work, the joint probability per electron of producing and detecting a photon is approximately: 
\[\alpha_\mathrm{\gamma}\cdot p_\mathrm{coh} = 10 ^{-5}\]  

The probability that is contributed by incoherent processes is unknown, but we consider it to be negligible in the context of this work. This is suggested by results presented in \cite{stoger-pollachFundamentalsCathodoluminescenceSTEM2019}: A measurement of CL, over a spectral range between 400 nm to 900 nm, in a thick (700 nm) silicon sample decreases by a factor of approximately 25 in intensity as the accelerating voltage is decreased from 200 keV to 40 keV. 

The intensity of coherent CL depends on the energy of the exciting electron, while the intensity of incoherent CL is typically proportional to the deposited energy. The amount of \textit{deposited} energy per electron actually \textit{increases} when lowering the electron energy from 200 keV to 40 keV, (stopping power for silicon is $\approx 0.5 \;\mathrm{keV/\upmu m}$ at 200 keV and $\approx 1.4\;\mathrm{keV/\upmu m}$  at 40 keV \cite{berger2017stopping}), this suggests that most of the signal at 200 keV stems from coherent radiation. 

\subsection{TEM instrumentation}\label{sec:si_TEM_instrumentation}
The experiments were carried out at the University Service Center for Transmission Electron Microscopy (USTEM) at TU Wien, using a FEI/TFS Tecnai G2 F20 TEM. The device uses a field-emission source, resulting in an energy spread of 0.9 eV (FWHM) at an acceleration voltage of 200 keV. 

All experiments were conducted in low-angle diffraction mode with an effective camera length of 380 m. The spot size and gun lens settings were adjusted to produce a current of $\approx3.1\;\mathrm{pA}$ as measured by the Timepix3-based detector (Advascope ePix) used for the experiments. This camera was retrofitted to the GIF Tridiem TV-port flange. The size of the beam on the sample plane was measured to be $54\;\mathrm{\mu m}$ by comparing it to a calibration target under identical imaging conditions directly after concluding the experiments (Ted Pella Inc., 2000 l/mm Cross Line Grating Replica, Prod. No. 677).

\subsection{Photon detection}\label{sec:si_photon_detection} 

Photons were collected using the Vulcan sample holder by Gatan. The first output port of the holder, which guides out the light collected by the bottom mirror, was connected to the input of the filter mount (Thorlabs FOFMF), using a standard multimode fiber cable (Thorlabs M76L02, compatible with the Vulcan holder's built-in fiber). The second output port, which guides light collected by the top mirror, was kept closed.  
The output of the filter mount was connected to the Picoquant PMA Hybrid 40 mod single photon detection module (SPDM) using another fiber cable (Thorlabs M74L02). The Picoquant PMA Hybrid 40 mod has a peak quantum efficiency of approximately $45 \%$ at $500\;\mathrm{nm}$, with a timing resolution of about 120 ps. This signal is connected to one of the input ports of the Swabian Instruments Time Tagger Ultra, which is used for timing the photon events. 
In order to filter the photon wavelength, we inserted band pass filters into the aforementioned filter mount (Thorlabs FBH 400-40, FBH 550-40 and FBH 650-40).   

Synchronisation was ensured by feeding the clock-out signal from the Advascope ePix (sensor: silicon, $100\;\mathrm{\mu m}$ thickness) into the time tagger as a reference clock. We used the time tagger's built-in softwareclock feature, which readjusts the recorded photon time tags according to the frequency standard provided by the reference signal. To fix a consistent starting time for photon time tags and electron time tags, we fed the trigger-out signal from the Advascope ePix into the time tagger and subtracted the time at which the trigger was registered from all photon time tags. 

In the first step, the microscope was aligned for the beam parameters used in the experiment (extraction, spotsize, gun lens, apertures, etc.). The size of the beam on the sample was also kept constant throughout the experiment. 
In order to align the focal point of the Vulcan holder's bottom mirror, the position of the sample holder was adjusted to yield the maximum photon count rate, as measured via the SPDM and time tagger. Contamination on the sample can markedly increase the photon count rate and thus negatively impact this alignment. Therefore, the illuminated area was inspected in TEM imaging and adjusted to avoid any visible contamination, while maintaining the highest possible photon count.   

\subsection{Data Processing}\label{sec:si_data_processing} 

The direct electron detector outputs a series of data packages containing the position (i.e. the pixel index), time of arrival (ToA) and time over the threshold (ToT) of each detection event \cite{poikelaTimepix365KChannel2014}. For each package received by the control PC, the photon time tags recorded over the same period of time are retrieved from the time tagger and saved together with the electron data. The time of arrival is digitized into 1.56 ns time bins. 

A few pixels, which are known to be defective are excluded. Additionally, a known timing issue introduces a fixed shift in ToA for each column of pixels. As the offset is always the same, this is corrected by subtracting the appropriate value for the respective pixel from the measured ToA.

Each electron hitting the detector triggers multiple detection events, mostly in adjacent pixels (2.8 events per electron on average at the settings used for the experiment). The clustering algorithm DBSCAN (using the implementation provided by the scikit-learn library, using eps = 3, and 50 ns as one unit of time) was used to extract the position and arrival time of each electron from their associated detection events. Within a cluster, the earliest arrival time and the position of the bottom left corner of each cluster (minimum x and y values) are chosen as representative of the arrival time and position of the electron. Clusters that show a total ToT below $750\;\mathrm{ns}$ are removed, as the measured ToA value is known to be less accurate for low energy (and therefore low ToT) events. 

The temporal cross-correlation histogram shown in Fig. \ref{fig:Setup}(c) is computed by evaluating: 
\begin{equation}
\{\Delta t_\mathrm{e\gamma}\}_\mathrm{Histogram} = \{t_\mathrm{e}-t_\mathrm{\gamma}|\; (|t_\mathrm{e}-t_\mathrm{\gamma}|\leq 200 \;\mathrm{ns})\}
\end{equation} 
The resulting time differences are then histogrammed using 1.5625 ns time bins, consistent with the cycle time of the Timepix' fast clock. In order to allow for more efficient processing, the electron and photon data is sorted by arrival time and split into smaller time intervals to be evaluated consecutively. When evaluating the time difference of events within a fixed time interval, boundary effects can skew the result, e.g. for a photon that arrives within the first $50\;\mathrm{ns}$ of the interval, it is impossible to find an electron that arrived $>50 \;\mathrm{ns}$ earlier while still being in the same interval. In order to avoid this bias, the first and last $200\;\mathrm{ns}$ for the photons in each interval are excluded. 

The coincidence signals shown in Figs.~\ref{fig:Setup}, \ref{fig:EFTEM} and \ref{fig:Photon_filtered} are computed as described in the main text: For each photon detection event (at $t_\mathrm{\gamma}$) we find the electron detection event (at $t_\mathrm{e}$) that matches the expected time delay $\mathbb{E}[\Delta t_\mathrm{e\gamma}]= \mathbb{E}[t_\mathrm{e}-t_\mathrm{\gamma}]$ most closely. This expected delay, which is due to differing path lengths and rise times of the detectors, is given by the maximum of the temporal cross-correlation histogram ($\approx -80\;\mathrm{ns}$). 

The cross-correlation histogram also clearly illustrates that we miss-classify a certain portion of uncorrelated events as correlated events. This is due to the Poissonian process governing both electron emission from the gun and the emission of uncorrelated photons as well as dark counts of the photon detector. As we understand these events to be uncorrelated, we can assume that their distribution will be the same for any time delay. Subtracting this background signal from the total coincidence signal will yield a result that better represents the distribution of correlated events. 

We sample the distribution of the uncorrelated background coincidences by evaluating the coincidence signal on the same measured data, for an interval with a large negative offset in the expected time delay ($[(\mathbb{E}[\Delta t_\mathrm{e\gamma}]-100\;\mathrm{ns})-\tau/2,(\mathbb{E}[\Delta t_\mathrm{e\gamma}]-100\;\mathrm{ns})+\tau/2]$ with $\tau = 50\;\mathrm{ns}$). For electron-photon pairs that have a time difference in this range, the electron arrived much later than its partner photon, which does not hold for correlated pairs.

\subsection{Measurement Procedure}\label{sec:si_measurement_procedure}

One of the main difficulties in the experiments was a drift of the electron energy spectrometer. For most of these measurements, data was acquired for several minutes. Inspecting the resulting data, it was not uncommon to see the selected energy range shifting by $1\;\mathrm{eV}$ or even more, thereby invalidating the measurement. In the measured energy range, close to the zero-loss peak, any shift in the transmitted energy window is accompanied by a significant change in electron flux. Therefore the rate of detected electrons was used as a figure merit to validate the stability of the spectrometer during the measurement. The measurement procedure consisted of: 
\begin{enumerate}
    \item Centering the energy slit on the zero-loss peak, resulting in the maximum electron rate (or placing the zero-loss peak at the edge for measurements with larger energy windows).
    \item Setting the correct energy offset for the required measurement by adjusting the acceleration voltage.
    \item Noting down the resulting electron rate. 
    \item Recording the data. 
    \item Verifying that the electron rate stayed close to the initially determined value throughout the measurement. 
\end{enumerate}

Additionally, a drift correction was applied to the recorded data by evaluating their center of mass in consecutive $50\;\mathrm{s}$ intervals.

\subsection{Evaluating the Accuracy of the Coincidence Matching}\label{numbers-for-accuracy-of-coincidence-matching}

In this work, we find coincident electron-photon pairs to identify those electrons that produced a photon and can therefore be correlated in momentum with that photon.  We consider two figures of merit in order to evaluate the accuracy of this matching procedure:     
\begin{enumerate}
    \item \textbf{Coincidence to Accident Ratio (CAR)}: This quantity gives the ratio of true vs.~false coincidence counts that contribute to the final signal.
This  quantity is given by:  \[\mathrm{CAR}= \frac{R_\mathrm{true}}{R_\mathrm{false}}\] with: 
    \begin{itemize}
        \item \(R_\mathrm{coin}\) \ldots{} the total rate of coincidence events, evaluated as described above. 
        \item \(R_\mathrm{false}\) \ldots{} the rate of background events, rate of false coincidences, evaluated with a large negative offset in time to only include uncorrelated pairs of events, as described above. 
        \item \(R_\mathrm{true} = R_\mathrm{coin}- R_\mathrm{false}\) \ldots{} the estimated rate of true coincidence events 
    \end{itemize}
    \item\textbf{Enhancement Compared to Energy Filtering ($A$)}: This quantity describes how much larger the fraction of photon-emitting electrons is in the coincidence-filtered distribution, compared to just energy filtering. 
    The fraction of photon-emitting electrons in the coincidence-filtered distribution is given by:
    \[P(\mathrm{photon\text{-}emitting|coincidence}) = \frac{R_\mathrm{true}}{R_\mathrm{coin}} = \frac{\mathrm{CAR}}{\mathrm{CAR}+1}\]
    The fraction of photon-emitting electrons in the energy-filtered signal is given by:
    \[P(\mathrm{photon\text{-}emitting|EFTEM}) = \frac{i\alpha_\mathrm{e}p_\mathrm{coh}}{R_\mathrm{EFTEM}}\]
    With $p_\mathrm{coh}$ being evaluated by integrating $\rho_\mathrm{TR}$ in the observed energy range. $R_\mathrm{EFTEM}$ is the rate of electrons measured in the energy-filtered signal, including only hits that correspond to a momentum transfer above the light line [i.e., inside the red circle shown in Fig.~\ref{fig:EFTEM}(a)]. 
    To evaluate this, we use the simulated values for the underlying probability $\rho_\mathrm{TR}$ and assume incoherent CL to be negligible in this situation, as justified in the section "Contributions from Incoherent CL". For $\alpha_\mathrm{e}$ we assumed a value of 0.26 to account for elastic scattering and the possibility to emit a bulk plasmon in addition to a TR photon. The enhancement factor considered is given by: 
    \[A = \frac{P(\mathrm{photon\text{-}emitting|coincidence})}{P(\mathrm{photon\text{-}emitting|EFTEM})}\]
\end{enumerate}

For the best measurement setting (energy-filtered TEM, energy range $2.5\;\mathrm{eV}$ to $3.5\;\mathrm{eV}$, no photon filter), we obtained an enhancement factor of $17.1$, with a CAR value of $28.7$.  

\end{document}